\documentclass[runningheads]{llncs}
\usepackage{epsfig}

\usepackage{graphicx} % Allows including images
\usepackage{wrapfig}
\usepackage{cutwin}
\usepackage{animate}
\usepackage{booktabs} % Allows the use of \toprule, \midrule and \bottomrule in tables
\usepackage[T1]{fontenc}    % use 8-bit T1 fonts
\usepackage{hyperref}       % hyperlinks
\usepackage{url}            % simple URL typesetting
\usepackage{booktabs}       % professional-quality tables
\usepackage{amsfonts}       % blackboard math symbols
\usepackage{nicefrac}       % compact symbols for 1/2, etc.
\usepackage{microtype}      % microtypography
\usepackage{bm}
\usepackage{listings}
\usepackage{multirow}
\usepackage{caption}
\usepackage{subfig}
\usepackage{amsmath}
\usepackage{color}
\usepackage{xcolor}
\usepackage{colortbl}
\usepackage{amssymb}
\usepackage{mathtools}
\usepackage{pifont}
\usepackage{float}
\usepackage{verbatim}
\usepackage{lipsum}

\usepackage[width=122mm,left=12mm,paperwidth=146mm,height=193mm,top=12mm,paperheight=217mm]{geometry}

\newcommand{\E}{\mathbb{E}}

\newcommand{\vx}{{\bm x}}

\begin{document}
	% \renewcommand\thelinenumber{\color[rgb]{0.2,0.5,0.8}\normalfont\sffamily\scriptsize\arabic{linenumber}\color[rgb]{0,0,0}}
	% \renewcommand\makeLineNumber {\hss\thelinenumber\ \hspace{6mm} \rlap{\hskip\textwidth\ \hspace{6.5mm}\thelinenumber}}
	% \linenumbers
	\pagestyle{headings}
	\mainmatter
	\def\ECCVSubNumber{4993}  % Insert your submission number here
	
	\title{Perceptual Image Super-Resolution with Progressive Adversarial Network} % Replace with your title
	
	% INITIAL SUBMISSION 
	\begin{comment}
	\titlerunning{ECCV-20 submission ID \ECCVSubNumber} 
	\authorrunning{ECCV-20 submission ID \ECCVSubNumber} 
	\author{Anonymous ECCV submission}
	\institute{Paper ID \ECCVSubNumber}
	\email{\{abc,lncs\}@uni-heidelberg.de}}
	\end{comment}
	%******************
	
	% CAMERA READY SUBMISSION
	%\begin{comment}
	\titlerunning{SR with PAN}
	% If the paper title is too long for the running head, you can set
	% an abbreviated paper title here
	%
	\author{Lone Wong \and
		Deli Zhao \and
		Shaohua Wan \and
		Bo Zhang}
	\authorrunning{Lone Wong, Deli Zhao et al.}
	% First names are abbreviated in the running head.
	% If there are more than two authors, 'et al.' is used.
	%
	\institute{Xiaomi AI Lab \\
		\email{\{stardust94605, zhaodeli\}@gmail.com}  \email{\{wanshaohua, zhangbo\}@xiaomi.com}}
	
	%\end{comment}
	%******************
	\maketitle
	
	\begin{abstract}
		Single Image Super-Resolution (SISR) aims to improve resolution of small-size low-quality image from a single one. With popularity of consumer electronics in our daily life, this topic has become more and more attractive.
		In this paper, we argue that the curse of dimensionality is the underlying reason of limiting the performance of state-of-the-art algorithms. To address this issue, we propose Progressive Adversarial Network (PAN) that is capable of coping with this difficulty for domain-specific image super-resolution. The key principle of PAN is that we do not apply any distance-based reconstruction errors as the loss to be optimized, thus free from the restriction of the curse of dimensionality. To maintain faithful reconstruction precision, we resort to U-Net and progressive growing of neural architecture. The low-level features in encoder can be transferred into decoder to enhance textural details with U-Net.  Progressive growing enhances image resolution gradually, thereby preserving precision of recovered image. Moreover, to obtain high-fidelity outputs, we leverage the framework of the powerful StyleGAN to perform adversarial learning. Without the curse of dimensionality, our model can super-resolve large-size images  with remarkable photo-realistic details and few distortions. Extensive experiments demonstrate the superiority of our algorithm over state-of-the-arts both quantitatively and qualitatively. 
		
	\end{abstract}

	\section{Introduction}
	Single image super-resolution (SISR)~\cite{SISR,ntire2017,ntire2018,ntire2019,deepsr} has been an active topic for decades because of its practical value in improving the visual quality of digital images. SISR addresses the ill-posed problem of estimating a super-resolution (SR) image from its low-resolution (LR) counterpart. Benefiting from the success of deep learning, the performance of SISR has been consistently improved by carefully designed architectures~\cite{SRCNN,FSRCNN,ESPCN,VDSR,LapSRN,EDSR,CARN,RCAN,RDN,towardRwSR} and diverse loss functions~\cite{perceptual,SRGAN,enhancenet,ESRGAN,zoomToLearn}. Despite those recent progress, the gap between papers and practical applications is still huge. Although these methods work well for common benchmark datasets, they often fail real-world images that are usually captured in the wild. Recent research attributes this to the lack of natural LR and SR image pairs and develops some different ways to acquire such image pairs~\cite{zoomToLearn,camLensSR,towardsRaw}. Even so, the diversity of real-world images and the complexity of image degradation are still far beyond our comprehension, which greatly complicates the mapping from low-resolution (LR) to high-resolution (HR) images. To this end, domain-specific super-resolution~\cite{fsrnet,SFTGAN,superFan,TDAE,yang2018hallucinating,yu2016ultra,lee2018attribute,SICNN,MTUN,learnDegradationFirst} simplifies this problem by targeting limited image domains. By leveraging the domain-specific information in training sets, this line of methods greatly improves generalization and robustness of image super-resolution, especially for large-scale super-resolution.
	
	However, image super-resolution suffers from a fundamental problem, say, the curse of dimensionality describing that the distance between two arbitrary data points  becomes indiscernible when the dimension is sufficiently high~\cite{NN99,Aggarwal01}. When measuring the distance between super-solved images and ground-truth images, the norm-based distances (e.g. $L_1$ and $L_2$ norms) are indispensable for existing algorithmic frameworks. For image super-resolution, however, the resulting dimension is very high. For example, the dimension is 1e+6 for an image of size $1000\times 1000$. Therefore,  the difficulty of image super-resolution from the curse of dimensionality ubiquitously exists in current algorithms as long as the distance-based reconstruction errors are applied.

	Hopefully, GAN-based models~\cite{goodfellow2014GAN} are capable of recovering visually plausible images. In particular, the recent style-based generator architecture for GAN algorithm (StyleGAN)~\cite{styleGAN,styleGAN2} can generate large-scale photo-realistic images from a low-dimensional random vector. But GAN-based models  usually suffer from many artifacts. Instead, PSNR-based models produce more faithful images but not visually plausible. And both situations get worse as the SR scale factor increases. To leverage the strength of GAN models, we propose a novel distance-free algorithm for domain-specific image super-resolution. Our main contributions are summarized as follows: \vspace{-0.1cm}
	\begin{itemize}
		\item To deal with the curse of dimensionality, we apply the progressive growing of neural architecture with skip connections instead of distance metrics. The semantic contents of high-resolution images are progressively derived and faithfully guaranteed from the low-resolution ones by fade-in layer. In this way, we remove the norm-based pixel losses such as $L_1$ and $L_2$, which helps us obtain high-quality reconstructions with high fidelity to source images.
		\item Adversarial learning is harnessed to enforce the outputs of the network to be highly photo-realistic. Different from the exiting SR methods, we only use the GAN loss. There is not any distance-based pixel losses involved in our framework. Therefore, our algorithm does not suffer from artifacts like blurriness and checkerboard patterns.
		\item Our distance-free framework may put a dent in not only image super-resolution, but also image denoising, image deblurring, etc. \vspace{-0.1cm}
	\end{itemize}
	To the best of our knowledge, our algorithm is the first that can produce the perceptually faithful $8\times$ high-quality super-resolution images without using distance-based reconstruction losses. 
	Particularly, our algorithm prefers the \textit{perceptually} accurate reconstruction of ground truth rather than pixel-wise precision. To highlight the critical difference from the conventional fashion of image super-resolution, therefore, we use the concept of \textit{perceptual} image super-resolution, as proposed in~\cite{PIRM2018}. From the practical point of view, we think that perceptual image super-resolution is more plausible and sound for the image super-resolution task. The characteristics of perceptual image super-resolution  will be illustrated in detail in section~\ref{se:experiment}.
	
	\section{Related Work}
	\textbf{Deep CNN for Super-Resolution}. 
	As a pioneer deep-learning-based SR method, Super-Resolution Convolutional Neural Network (SRCNN) ~\cite{SRCNN} outperforms traditional algorithms with a deep convolutional neural network. Different from taking upsampled images as input,  Efficient Sub-Pixel Convolutional Neural network (ESPCN)~\cite{ESPCN} and Fast Super-Resolution Convolutional Neural Network (FSRCNN)~\cite{FSRCNN} upsample images at the end of the networks. The reduction in the number of large-scale operations makes them less time-consuming compared to SRCNN. To better harness the power of deep-learning models, Kim \textit{et al.} proposed Very Deep Super-Resolution (VDSR)~\cite{VDSR}, a deep network which introduces residual learning to ease training process and achieves remarkable improvement in accuracy. By removing unnecessary modules in conventional residual networks, Lim \textit{et al.}~\cite{EDSR} proposed Enhanced Deep Super-Resolution network (EDSR) and Multi-scale Deep Super-Resolution System (MDSR), which also achieve significant improvement.
	
	To improve the visual quality of SR results, perceptual loss~\cite{perceptual} is applied by minimizing the error in the feature space of a well-trained model instead of directly in the pixel space. Besides, contextual loss~\cite{contextual} is developed to generate images using an objective that focuses on the contextual similarity rather than merely spatial location of pixels. Ledig \textit{et al.} introduced Residual Neural Network (ResNet)~\cite{Resnet} to construct a deeper network named SRResNet~\cite{SRGAN}. In this work, they also proposed GAN for Image Super-Resolution (SRGAN) using perceptual loss and GAN loss to achieve photo-realistic SR. Following this approach, both SISR through automated texture synthesis (EnhanceNet)~\cite{enhancenet} and Enhanced Super-Resolution Generative Adversarial Networks (ESRGAN)~\cite{ESRGAN} use GAN-based models to attain visually plausible SR results. Recovering realistic texture in image super-resolution by deep Spatial Feature Transform (SFTGAN)~\cite{SFTGAN} finds that more photo-realistic results can be obtained with a correct categorical prior.
	
	Of all aforementioned literature, photo-realism is usually attained by adversarial training with GANs, but their predicted results may not be faithfully reconstructed and sometimes produce unpleasing artifacts. Just like what Blau \textit{et al.}~\cite{blau2018perception} have proved, the distortion and perceptual quality are at odds with each other.
	
	\textbf{Domain-Specific Super-Resolution}. 
	As for domain-specific SR, face SR a.k.a. face hallucination has been the most widely studied subject. These methods utilize facial priors explicitly or implicitly. Super-resolution of real-world low resolution faces in arbitrary poses with GANs (Super-FAN)~\cite{superFan} and Multi-Task Upsampling Network (MTUN)~\cite{MTUN} both apply facial landmarks from Face Attention Network (FAN)~\cite{FAN} to guarantee the consistency in end-to-end learning. And Face Super-Resolution Network (FSRNet)~\cite{fsrnet} tries not only facial landmark heatmaps but also face parsing maps as prior constraints. Super-Identity Convolutional Neural Network (SICNN)~\cite{SICNN} uses a super-identity loss function to recover the person identity. Transformative-Discriminative Autoencoder (TDAE)~\cite{TDAE} adopts a decoder-encoder-decoder framework to handle noisy face images. Despite the fact that they get better SR results for face, they have settled for a low-resolution image generation, greatly limiting the visual quality of SR images.

	\begin{figure}[t]
		\centering
		\includegraphics[scale=0.7]{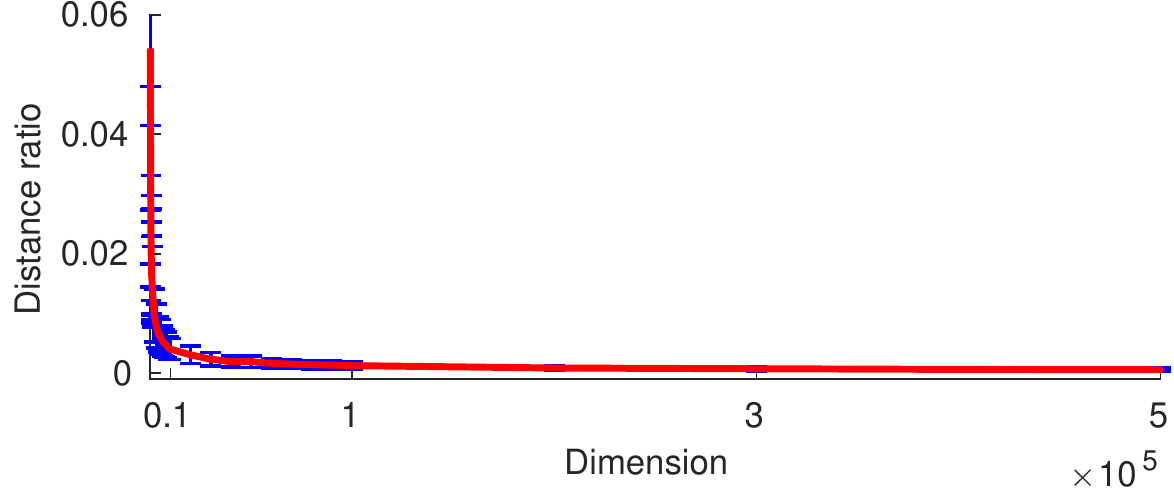}\vspace{-0.2cm}
		\caption{ Illustration of the curse of dimensionality. The distance ratio is exactly the formula in (\ref{eq:dimension}). The data set for each dimension consists of 500 data points randomly sampled from uniform distribution. We search $\log(500)$ nearest neighbors for each point. Among these nearest neighbors, the nearest one is taken as $\text{dist}_{\min}(\vx)$ and the farthest one as the $\text{dist}_{\max}(\vx)$.}
		\label{fig:dimension} \vspace{-0.3cm}
	\end{figure}
	\section{Curse of Dimensionality}\label{se:dimension}
	The principal argument about the issue in high-dimensional spaces is that the concept of distance-based nearest neighbors is no longer meaningful when the dimension goes sufficiently high~\cite{NN99,Aggarwal01}. This proposition is so important that it is worth being written rigorously, i.e., \vspace{-0.1cm}
	\begin{equation}\label{eq:dimension}
	\lim_{d_x\rightarrow \infty} \E \left [  \frac{\text{dist}_{\max}(\vx) - \text{dist}_{\min}(\vx)}{\text{dist}_{\min}(\vx)} \right ] \rightarrow 0,~~~
	\text{if} \lim_{d_x\rightarrow \infty} \text{var} \left  [\frac{\|\vx\|}{\E[\|\vx\|]} \right  ] \rightarrow 0,  \vspace{-0.1cm}
	\end{equation}
	where $\text{dist}_{\max}(\vx)$ and $\text{dist}_{\min}(\vx)$ represent the maximum distance and the minimum distance to $\vx$ in the given data set, respectively. The above limit implies that for sufficiently large $d_x$, the data points become \textit{spatially} indiscernible by norm-based distance measures in high-dimensional spaces. To illustrate this effect, we compute the ratio of distance discrepancy in formula (\ref{eq:dimension}) in various dimensions. Figure~\ref{fig:dimension} shows that the distance ratio is prone to converge and approach zero with negligible standard deviation after ten thousand dimensions.

	The curse of dimensionality suggests that we would run the risk of adopting the norm-based measures or analogous counterparts to quantize the discrepancy between super-resolution images, e.g. reconstruction error. 
	In our opinion, the subtle artifacts produced by networks for SR are partially due to this reason. From the view of dimensionality, we should avoid using distance measures like  $L_1$, $L_2$, and perceptual loss for very high-dimensional super-resolution task. In this paper, we propose an alternative solution to replace the distance measures in image super-resolution.

	\section{Progressive Adversarial Network}
	The proposed PAN aims to estimate a SR image from its LR counterpart, synthesizing photo-realistic and faithful high-quality image while preserving the consistency with the LR image in content. And simultaneously, it is expected that the algorithm is able to generalize well to unseen real-world data. To this end, we need to achieve the following two goals.  \vspace{-0.1cm}
	\begin{itemize}
		\item Control the reconstruction error without distance measures, as interpreted in section~\ref{se:dimension}. We fulfill this condition with progressive growing of a partial U-Net.   
		\item Maintain the high-quality photo-realistic effect. This goal is attained by adversarial learning. A GAN architecture with random noise injection is designed to synthesize high-quality images.  \vspace{-0.1cm}
	\end{itemize}
	The details are given in the following sub-sections.
	\begin{figure}[t]
		\centering
		\includegraphics[width=\linewidth]{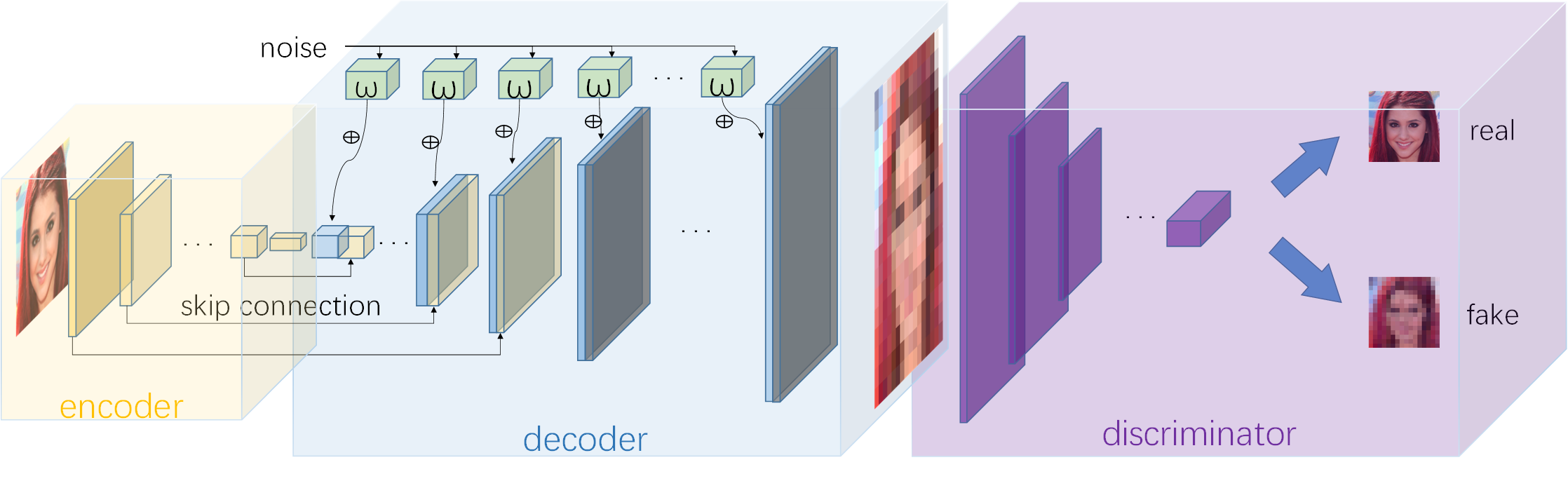} \vspace{-0.2cm}
		\caption{Algorithmic architecture. The network here takes an input image having resolution of $128\times128$ pixels and super-resolves it to $1024\times1024$. We concatenate feature maps from encoder (yellow) to decoder (blue) at low resolution, while at high resolution we substitute skip connection with random noise (gray). The discriminator learns incrementally to distinguish between real and fake images as the training proceeds, thus increasing the spatial resolution of generated images.}\vspace{-0.3cm}
		\label{fig0}
	\end{figure}
	\subsection{Progressive Super-Resolution}
	In our distance-free framework, we harness progressive growing of a partial U-Net to guarantee the faithful generation of resolved images. Progressive learning shows a good potential in ProGAN~\cite{progressiveGAN} and StyleGAN~\cite{styleGAN} and this idea is intuitively suitable for super-resolution task~\cite{progressiveSuperRes18}. The key role of progressive super-resolution is that the high-resolution image can be \textit{gradually} generated from the low-resolution one. In this way, the reconstruction accuracy can be well assured without the constraint of distance-based losses, thus freeing our algorithm from the curse of dimensionality. Formally, the process of progressive growing can expressed as:  \vspace{-0.1cm}
	%
	% \begin{equation}
	% \bm x \overset{\chi_1}{\longmapsto} \bm x^{\text{2x}}, ~
	% \bm x \overset{\chi_1}{\longmapsto} \bm x^{\text{2x}} \overset{\chi_2}{\longmapsto} \bm x^{\text{4x}}, ~
	% \bm x \overset{\chi_1}{\longmapsto} \bm x^{\text{2x}} \overset{\chi_2}{\longmapsto} \bm x^{\text{4x}} \overset{\chi_3}{\longmapsto} \bm x^{\text{8x}}, \vspace{-0.1cm}
	% \end{equation}
	%
	%
	\begin{equation}
	\left\{ \begin{aligned}
	& \bm x \overset{\chi_1}{\longmapsto} \bm x^{\text{2x}} \\
	& \bm x \overset{\chi_1}{\longmapsto} \bm x^{\text{2x}} \overset{\chi_2}{\longmapsto} \bm x^{\text{4x}} \\
	% & \cdots \\
	& \bm x \overset{\chi_1}{\longmapsto} \bm x^{\text{2x}} \overset{\chi_2}{\longmapsto} \bm x^{\text{4x}} \overset{\chi_3}{\longmapsto} \bm x^{\text{8x}}
	\end{aligned}
	\right.,  \vspace{-0.1cm}
	\end{equation}
	where $\bm x^{\text{2x}}$ means that the size of $\bm x^{\text{2x}}$ is $2\times$ larger than that of $\bm x$ and the other is the same, and $\chi_i$ is the corresponding generator. The parameter of $\chi_j$ ($j>i$) is learned with the pretrained $\chi_i$ for each resolution. 
	
	When the size of the output image is no more than that of the input one, U-Net~\cite{Unet} is employed as the generator, which contains the encoder-decoder parts and skip connections. These connections are useful to save insights from different abstraction levels and transfer them from the encoder to the decoder network. Otherwise, only progressive growing of image resolution is involved.  An overview of the proposed PAN is shown in Fig.~\ref{fig0}.
	
	To be specific, our progressive super-resolution is two-fold. First, the fade-in layer is introduced to ease the training on higher resolutions. As illustrated in Fig.~\ref{fig1}, the toRGB layer projects feature maps to images and the fromRGB layer performs the reverse mapping. The weight $\alpha$ increases linearly from 0 to 1 as the training proceeds, making the new layer fade in the network smoothly. Second, both input LR images and real images used to train discriminator are provided in a coarse-to-fine way. Namely our training starts with a resolution of $8\times8$ pixels. So the input image and the ground truth image should both be degraded to $8\times8$ pixels, which makes it much easier for generator to produce ``real'' images and succeed in fooling the discriminator.

	\subsection{Adversarial Learning without Distance Measures}\label{AA}
	To obtain the photo-realistic quality of generated SR images, the conventional way is to resort to adversarial learning. In general,  the existing algorithms follow the GAN framework of Pix2Pix~\cite{PIX2PIX} that firstly proposes to combine the adversarial loss with norm-based regularization loss, so that the generator is trained not only to fool the discriminator but also to generate images as close to ground-truth as possible. As we analyze in section~\ref{se:dimension}, the distance measures such as $L_1$ and $L_2$ norms lead to ambiguous artifacts due to the curse of dimensionality. Typically, norm-based reconstruction losses aim to achieve higher Peak Signal-to-Noise Ratio (PSNR), but usually leads to blurry images. In addition, it is a very tricky step to balance different losses, say adversarial loss, reconstruction loss, and perceptual loss. For our distance-free algorithm, however, we only need a GAN model that is capable of significantly enhancing the visual quality of SR images. We use non-saturating loss~\cite{goodfellow2014GAN} with ${R_1}$-regularization~\cite{regularizationR1} as loss function. The discriminator loss is defined as: \vspace{-0.1cm}
	\begin{equation}
	{L_d} =  - {\mathbb{E}_{\vx \sim {p_x}}}[\log (D(\vx))] - {\mathbb{E}_{\hat \vx \sim {p_{\hat x}}}}[\log (1 - D(\hat \vx))] + \gamma \underset{ {\vx}\sim p_x }{\mathbb{E}}[  \|\nabla_{{\vx}}D({\vx})\|^2_2 ],\vspace{-0.1cm}\label{eq0}
	\end{equation}
	where $\gamma$ is the hyper-parameter to weigh the ${R_1}$-regularization. The adversarial loss for generator is non-saturating loss as: \vspace{-0.1cm}
	\begin{equation}
	{L_g} =  - {\mathbb{E}_{\hat \vx \sim {p_{\hat x}}}}[\log (D(\hat \vx))].\vspace{-0.1cm}\label{eq1}
	\end{equation}

\begin{figure}[t]
		\centering
		\includegraphics[width=\linewidth]{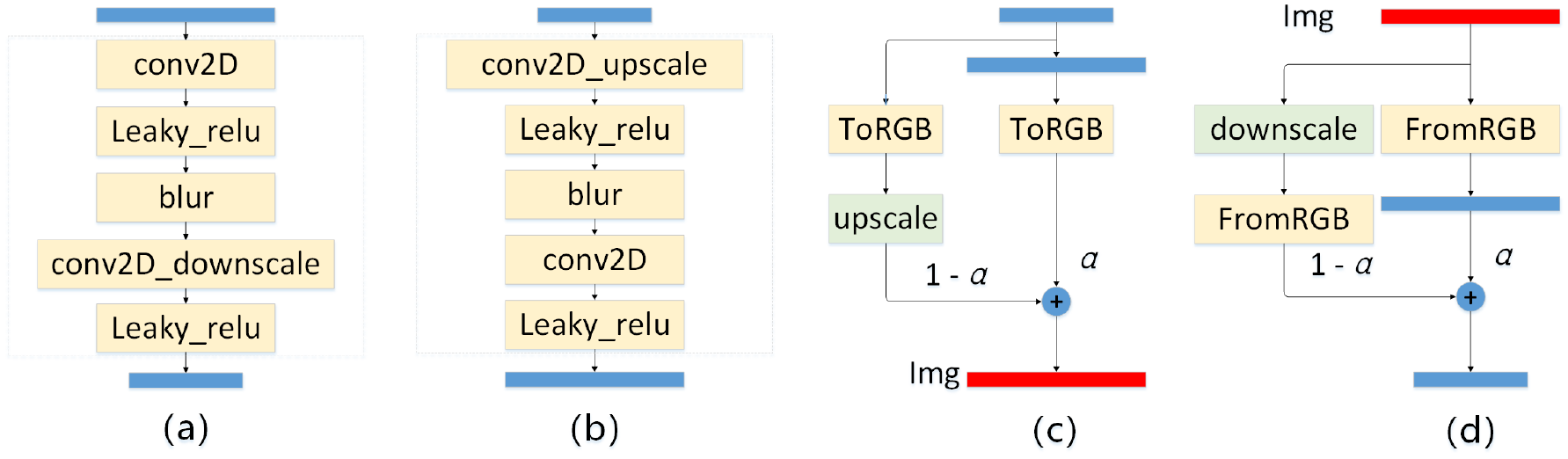}\vspace{-0.2cm}
		\caption{ Neural networks. (a) Block in encoder and discriminator.  (b) Block in decoder.  (c) Fade-in layer in generator. (d) Fade-in layer in discriminator. The $conv2D\_downscale$ layer represents conv2d layer with stride 2 while $conv2D\_upscale$ represents transpose of conv2D with stride 2.} \vspace{-0.3cm}
		\label{fig1}
	\end{figure}

	Following StyleGAN~\cite{styleGAN}, the structure of blocks to encode and decode features are shown in Fig.~\ref{fig1}. Encoder and discriminator are similar in structure and decoder and discriminator are mirror images of each other. 
	
	\subsection{Random Noise Injection}
	For StyleGAN~\cite{styleGAN}, the authors discover an appealing property of random noise in enhancing the quality of generator. The photo-realistic details of generated images can be significantly improved by injecting random noise into feature maps of generator, i.e. \vspace{-0.1cm}
	\begin{equation}
	\bm y_i \leftarrow  \bm y_i + w_i \bm \varepsilon, \vspace{-0.1cm} %
	\end{equation}
	where $\bm y_i$ is the $i$-th channel of feature maps,  $\bm \varepsilon$ is the random noise of same size with $\bm y_i$,  and $w_i$ the learnable parameter of scaling the noise. For our PAN architecture, we adopt this type of noise injection as well. The role of random noise mainly regularizes the deep neural network~\cite{Noh2017noise,You2018noise}, thus stabilizing the training process and facilitating the algorithmic convergence. 
	
	It is worth emphasizing that the advantage of noise injection cannot be achieved for GAN models with distance measures. $\bm \varepsilon$ is randomly sampled for each update during training. The different $\bm \varepsilon$ will result in the detail alteration for generated images. This operation amounts to randomly perturbing distance losses, thus hardening the convergence of the algorithm. However, our PAN framework is free from this negative influence and is able to take the advantage of noise injection, as StyleGAN does.
	
	\section{Experiment}\label{se:experiment}
	In this section, we first explain some training details and compare PAN with state-of-the-art SR methods on three commonly-used image quality metrics: PSNR, Structural Similarity Index (SSIM)~\cite{SSIM}, and Naturalness Image Quality Evaluator (NIQE)~\cite{NIQE}. As noted and shown in the previous work~\cite{Chen2018dark,Yu2018exposure,PIRM2018}, these pixel-based metrics sometimes cannot accurately reflect the visual quality of resulting images and are not suitable for perceptual image super-resolution. To remedy this, we also use another two metrics for GANs: Fr\'{e}chet Inception Distance (FID)~\cite{FID} and Sliced Wasserstein Distance (SWD)~\cite{progressiveGAN} to estimate realism by measuring how our SR results resemble real face images. 
	Then we conduct the ablation experiment to investigate how each part of the network influences the capability of performing super-resolution.
	
	Actually, these pixel-level metrics such as PSNR, SSIM, and NIQE are not suitable for perceptual image super-resolution we raise, because perceptual image super-resolution aim to recover large-size images that are perceived holistically accurate rather than pixel-wise precision. The results of these metrics are just for reference of reconstruction precision for our algorithm.
	
	\subsection{Implementation Details}
	As shown in Fig.~\ref{fig0}, our model super-resolves small images with a scaling factor of 8 (scaling images from $128\times128$ to $1024\times1024$). We start with $8\times8$ resolution and stabilize the network in 600k iterations at each resolution (8, 32, 64, 128, 256, 512, and 1024). Every time right after doubling the resolution, the new layer fades in smoothly and it takes 600k iterations to be completely integrated into the network. The entire network is trained in an end-to-end manner using loss function in Eq. \eqref{eq0} and Eq. \eqref{eq1} alternately with $\gamma=5$. Adam optimizer is used with $\beta_1=0$, $\beta_2=0.99$, $\epsilon$=1e-8 and learning rate for different resolutions follows this setting $\{8\sim64:0.001,  128:0.0015,  256:0.002, 512\sim1024:0.003\}$. We use different minibatch sizes according to $\{8:64, 16:32, 32:16, 64:8, 128\sim1024:4 \}$ to avoid out-of-memory problem. Our training set and test set are Flickr-Faces-HQ (FFHQ) dataset~\cite{styleGAN} and CelebA-HQ~\cite{progressiveGAN}, respectively. It takes about 5 days to complete the training process using 4 Tesla P40 GPUs.
	\setlength{\tabcolsep}{4pt}
	\begin{table}[t]
		\begin{center}
			\caption{Quantitative comparison with state-of-the-art methods. SWD $\times10^3$ is given for levels 512 and 256.}
			\label{table:metrics}
			\begin{tabular}{lllllll}
				\hline\noalign{\smallskip}
				Methods  & PSNR $\uparrow$ & SSIM $\uparrow$ & NIQE $\downarrow$& FID $\downarrow$ & \multicolumn{2}{l}{SWD $\times10^3$} $\downarrow$\\
				& & & & & 512 & 256   \\
				\noalign{\smallskip}
				\hline
				\noalign{\smallskip}
				Bicubic & 30.79 & 0.84 & 15.95 &76.53 &98.31 & 42.75   \\
				ESRGAN & 25.04 & 0.65 & 7.50 & 8.03 & 40.15 & 34.08  \\
				CARN & 31.54 & 0.86 & 14.73 & 27.31 & 133.39 &  37.23  \\
				RCAN & 31.33 & 0.86 & 14.31 & 20.47 & 135.68 & 38.01  \\
				PAN (ours) & 26.81 & 0.73 & 9.71 &\textbf{5.16} & \textbf{33.51} & \textbf{28.35}  \\
				\hline
			\end{tabular}
		\end{center}\vspace{-0.3cm}
	\end{table}
	\setlength{\tabcolsep}{1.4pt}
	
	\subsection{Quantitative Evaluation}
	We compare our models with bicubic scale and state-of-the-art SISR methods including ESRGAN~\cite{ESRGAN}, Residual Channel Attention Networks (RCAN)~\cite{RCAN} and Cascading Residual Network (CARN)~\cite{CARN}, on top 5000 images of CelebA-HQ. For the reason that $\times8$ scale is not supported in the official implementation of ESRGAN and CARN, our evaluation is performed on $\times4$ scale. Our model originally produces $\times8$ SR image. To compare with other methods, we apply a $2\times2$ average pooling to our results and make it a $\times4$ SR image. Quantitative comparisons are given in Table~\ref{table:metrics}.
	
\begin{figure}[t]
		\centering
		\includegraphics[width=\linewidth]{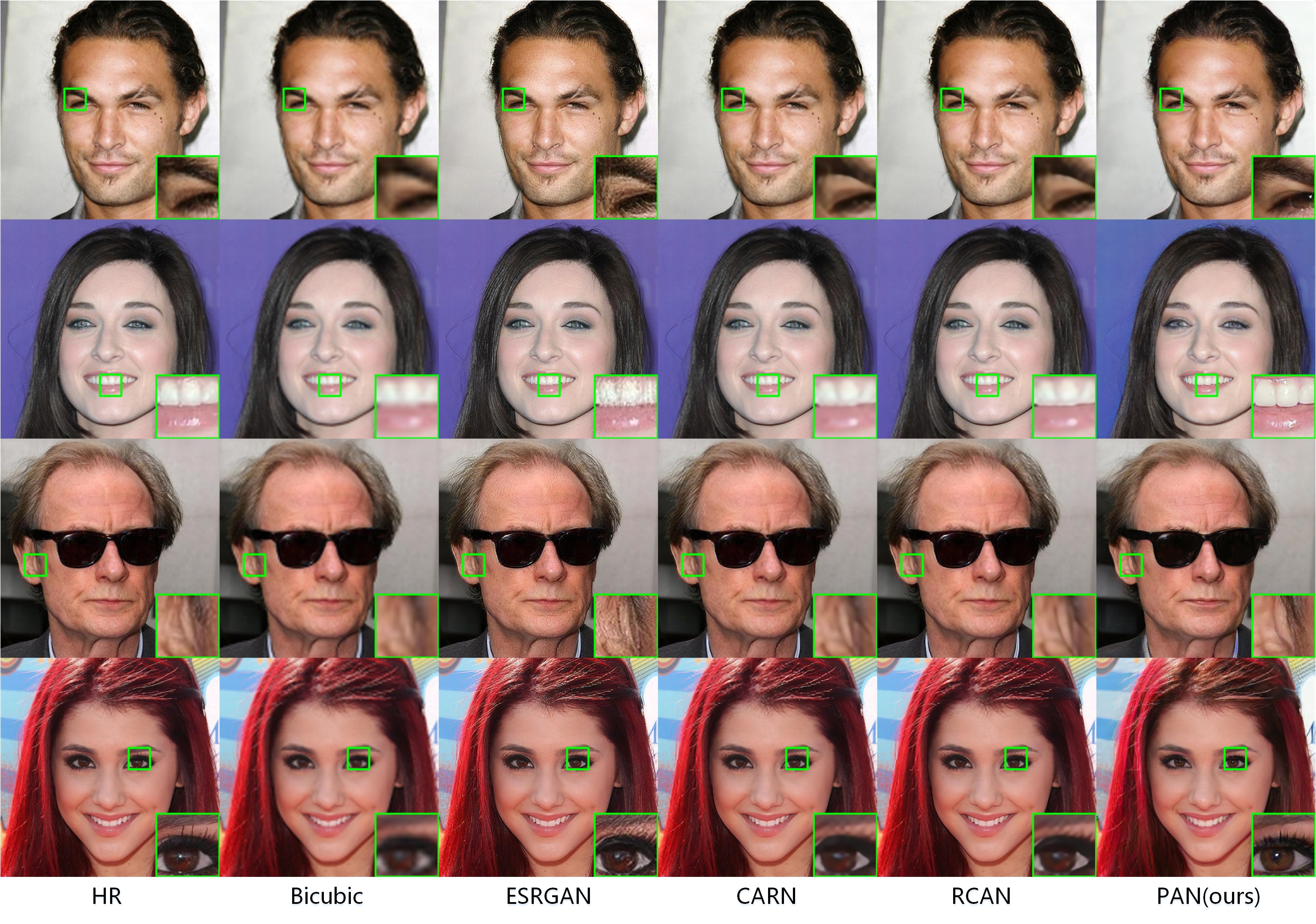}\vspace{-0.2cm}
		\caption{Super-resolution result of our methods compared with existing methods.} \vspace{-0.3cm}
		\label{fig3}
	\end{figure}

	Of all the compared methods, our model performs the best for FID and SWD, and second best for NIQE. Indeed, our method significantly reduces distortion and greatly improves perceptual quality as shown in Fig.~\ref{fig3}. We observe that ESRGAN tends to produce over-sharp image with ringing artifacts and RCAN, CARN cannot recover details and suffers from blurry appearance. In contrast, our PAN produces SR images much more photo-realistic than others, especially when we zoom in and examine image details. For example, ESRGAN fails to recover the teeth shape (the second row) and the eyelash style (the last row), and over-sharpen the textural details of the eye (the first row) and the ear (the third row). Instead, our PAN algorithm generates more perceptually faithful results with ground truth than ESRGAN, even though ESRGAN's PSNR, SSIM, and NIQE are better than ours. 
	
	What's more intriguing is that our method frequently generates SR results even visually better than original HR images, as the second face example shows in Fig.~\ref{fig3}. This unique strength stems from the perceptual characteristic of our algorithm without the constraint of pixel losses, which is beyond the capability of existing state-of-the-art algorithms that optimize distance-based reconstruction errors. These results solidly support the plausibility of perceptual image super-resolution.

\begin{figure}[t]
		%\vskip 0.2in
		\begin{center}
			\begin{tabular}{cc}
				\includegraphics[scale=0.15]{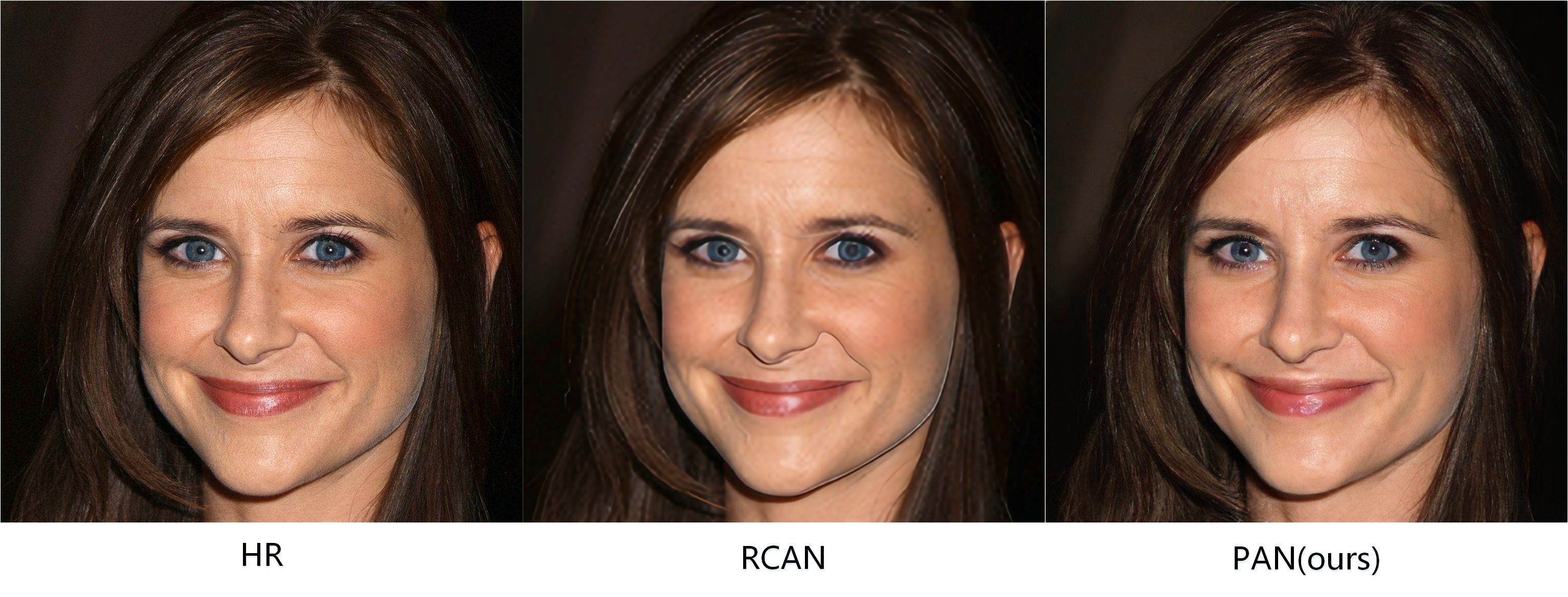}\hspace{0.cm}
				&
				\includegraphics[scale=0.47]{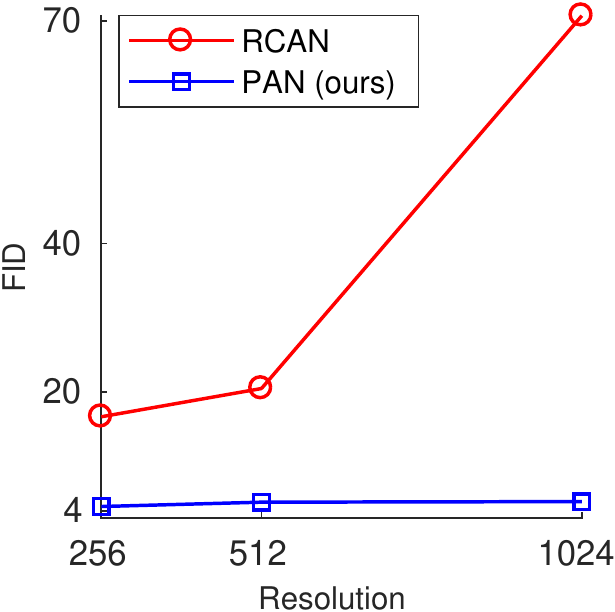}\\
				(a) Super-resolution images ($1024\times 1024$). &  (b) FID accuracy.  
			\end{tabular}
		\end{center} \vspace{-0.5cm} 
		\caption{$8\times$ SR comparison.  } \vspace{-0.1cm}
		\label{fig_8xsr}
	\end{figure}

	To further demonstrate the power of our method, we compare our results with RCAN\footnote{Among compared state-of-the-art methods, only RCAN's official implementation directly supports $\times8$ super-resolution. So, we only compare RCAN in this experiment.} on different SR scales ($\times2$, $\times4$, $\times8$). As shown in Fig.~\ref{fig_8xsr}(b), when we increase the scale factor, the FID of RCAN results dramatically increases, which means a significant degradation in image quality. We attribute this to the high-dimensional output at end of the network which typically has a large tile size. As a comparison, our algorithm is quite stable at different scales and consequently gives a much better SR result at $\times8$ resolution, implying that PAN is rather robust to dimension varying.
	Images in Fig.~\ref{fig_8xsr}(a) also confirm the quantitative superiority of PAN. The obvious quality difference can be revealed by zooming in the facial details.

	More results are shown at \url{https://lonew.github.io/pan-sr/}.
	\subsection{GAN Loss}
\begin{figure}[t]\vspace{-0.2cm}
		\centering
		\includegraphics[width=\linewidth]{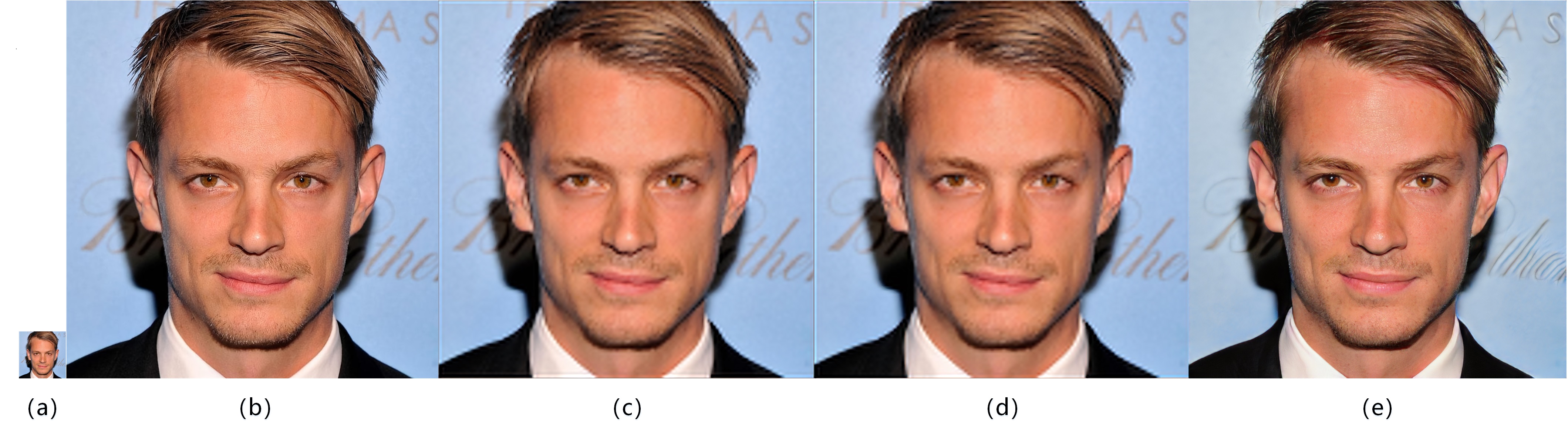}\vspace{-0.2cm}
		\caption{SR results with different loss functions. (a) Input image with $128\times128$ pixels. (b) HR image with $1024\times1024$ pixels. (c) SR image using $L_1$ loss. (d) SR image using $L_2$ loss. (e) SR image using GAN loss.} \vspace{-0.3cm}
		\label{fig_ganloss}
	\end{figure}

	To verify the advantage of GAN loss, we compare it with $L_1$ and $L_2$ losses. This evaluation is performed on $\times8$ SR results. Quantitative comparisons are given in Table~\ref{table:metrics_8x}. The model with $L_1$ or $L_2$ loss has better PSNR and SSIM, but generates blurry images with zipper artifacts, while the model with GAN loss produces clear and sharp images with better details. Though these tiny details such as hair and wrinkle are not identical to original HR image, the overall appearance visually stays the same and it is really hard to distinguish the difference as shown in Fig.~\ref{fig_ganloss}. The difference of tiny details is caused by random noise injection. However, it is partially due to random noise to make recovered SR image more holistically photo-realistic.
	\vspace{-0.2cm}
	\setlength{\tabcolsep}{4pt}
	\begin{table}[t]
		\begin{center}
			\caption{Quantitative comparison with different test methods, SWD $\times10^3$ is given for levels 1024, 512}
			\label{table:metrics_8x}
			\begin{tabular}{lllllll}
				\hline\noalign{\smallskip}
				Methods  & PSNR$\uparrow$ & SSIM$\uparrow$ & NIQE$\downarrow$ & FID$\downarrow$ & \multicolumn{2}{l}{SWD $\times10^3$}$\downarrow$\\
				& & & & & 1024 & 512  \\
				\noalign{\smallskip}
				\hline
				\noalign{\smallskip}
				$L_1$ loss & 28.53 & 0.76 & 14.76 &88.28 &111.33 &  73.94\\
				$L_2$ loss & 28.19 & 0.77 & 14.91  &89.17 &134.48 & 77.98 \\
				non-progressive & 24.17 & 0.59 & 10.34 &8.66 & 187.37 & 36.77\\
				non-noise & 25.29 & 0.66 & 12.02 &5.01 & 36.08 & 29.97  \\
				PAN & 25.88 & 0.65 & 10.07 & 5.24& 82.78 & 31.75  \\
				\hline
			\end{tabular}
		\end{center}\vspace{-0.3cm}
	\end{table}
	\setlength{\tabcolsep}{1.4pt}
	\begin{figure}[t]
		\centering
		\includegraphics[width=\linewidth]{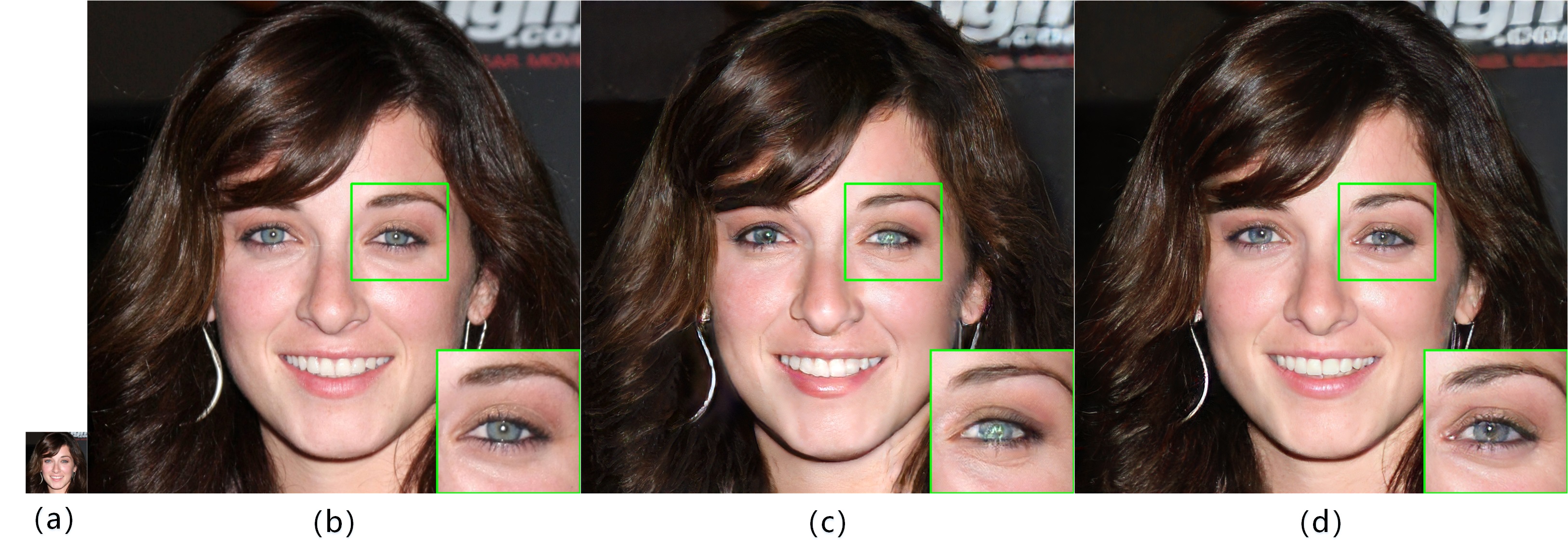}\vspace{-0.2cm}
		\caption{Progressive learning: (a) Input image with $128\times128$ pixels. (b) HR image with $1024\times1024$ pixels. (c) SR image without progressive learning. (d) SR image with progressive learning.} \vspace{-0.3cm}
		\label{fig_prog}
	\end{figure}
	\subsection{Progressive Learning}
	Intuitively, progressive learning makes training process more stable and thus helps converge to a considerably better optimum. To verify this, we train our PAN algorithm without progressive learning. Table~\ref{table:metrics_8x} illustrates the effectiveness of progressive learning quantitatively. There is an obvious degradation for SR images from the non-progressively trained network. In Fig.~\ref{fig_prog}, we observe that the progressive version of PAN largely improves the visual quality and reduces the distortion in image details. 
	\begin{figure}[t]
		\centering
		\includegraphics[width=\linewidth]{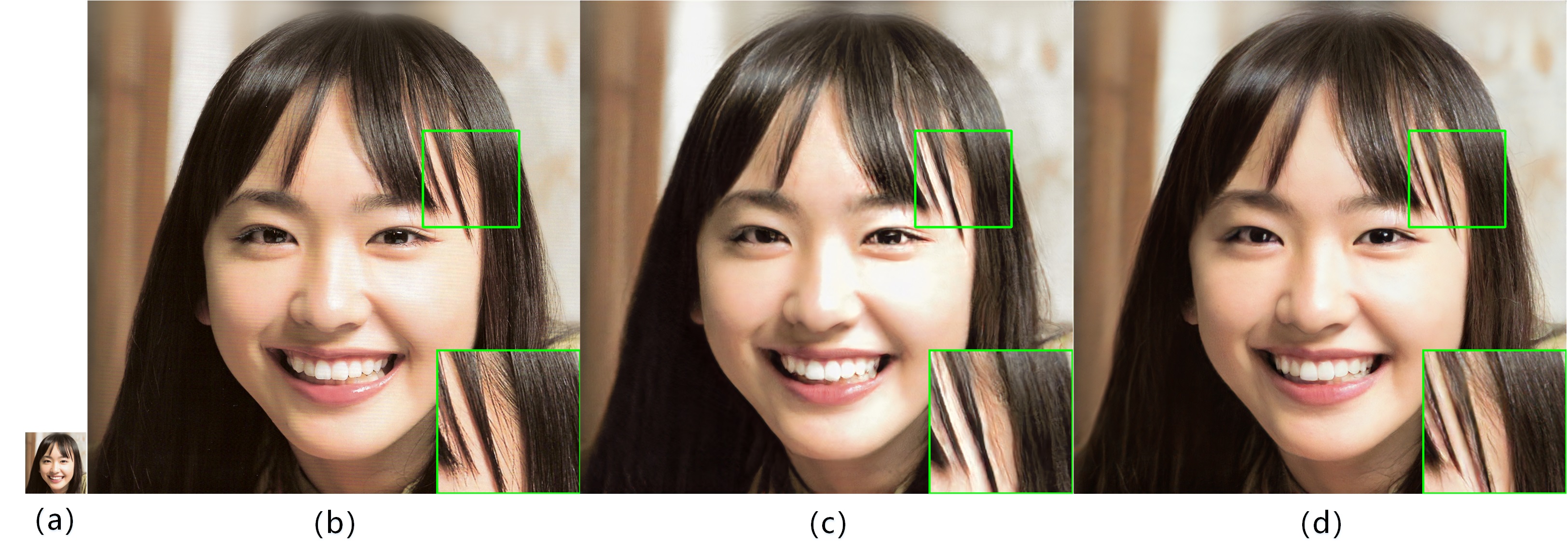}\vspace{-0.2cm}
		\caption{Random Noise Injection. (a) Input image with $128\times128$ pixels. (b) HR image with $1024\times1024$ pixels. (c) SR image without noise injection. (d) SR image with noise injection.} \vspace{-0.3cm}
		\label{fig_noise}
	\end{figure}
	\subsection{Random Noise Injection}
	StyleGAN indicates the fact that random noise leads to small stochastic variation in generated images. In our networks, we add random noise to feature maps at each resolution. Instead of rigidly following the distribution of LR image, noise injection allows the generator to produce more  variation in image details, which helps generate more photo-realistic and texture-rich images to fool the discriminator. Wondering what it would be like without these noise, we train a non-noise version of PAN. Table~\ref{table:metrics_8x} shows that noise injection only improves the NIQE metric and on the contrary, it makes FID and SWD drop a little. But in fact, these noise does have a huge influence in making the SR images more natural and realistic without changing holistic appearance, as shown in Fig.~\ref{fig_noise}. 
	\begin{figure}[t]
		\centering
		\includegraphics[width=\linewidth]{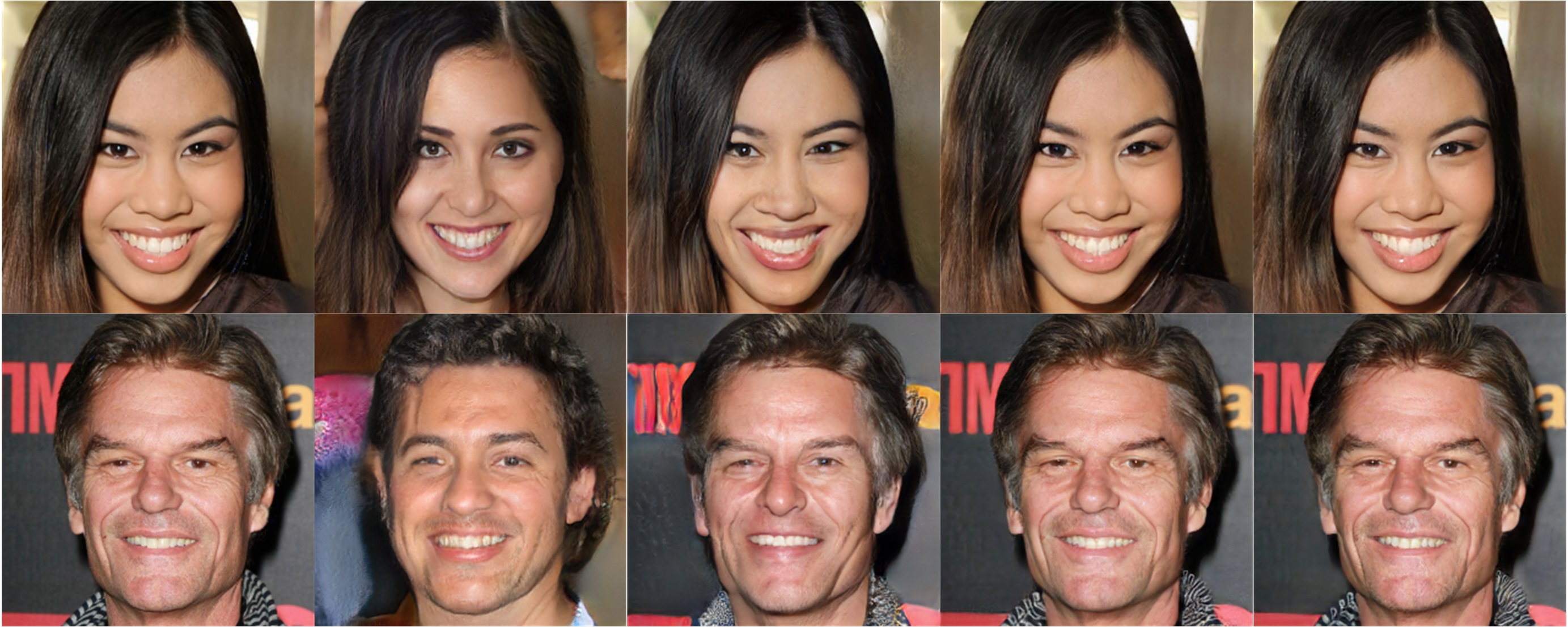}\vspace{-0.2cm}
		\caption{Skip connections. Images on the left are ground-truth HR images. The others are SR images with different levels of skip connections.}\vspace{-0.3cm}
		\label{fig4}
	\end{figure}
	\begin{figure}[t]
		\centering
		\includegraphics[width=\linewidth]{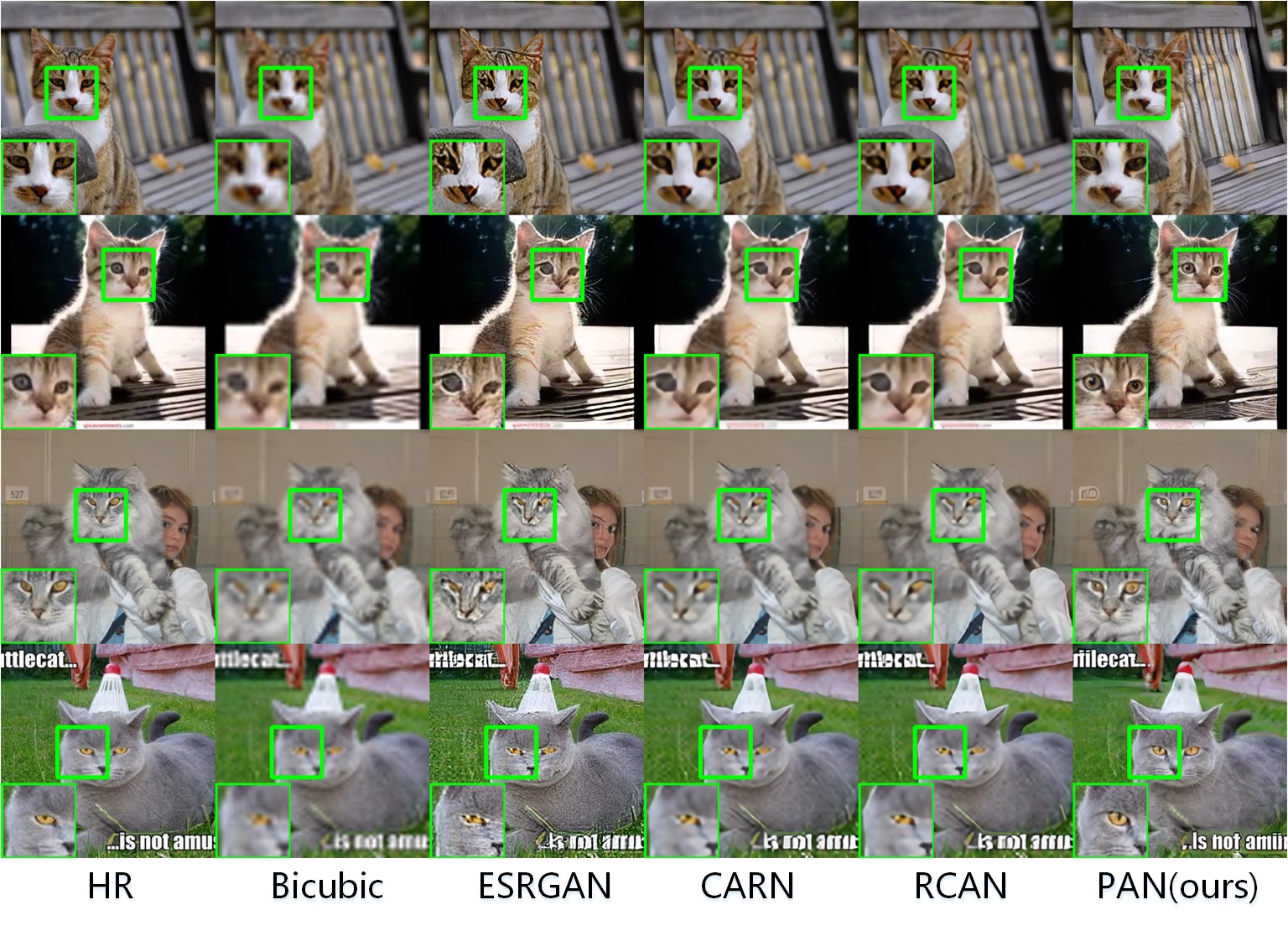}\vspace{-0.2cm}
		\caption{Visual results for different methods($\times4$ SR) on LSUN cat dataset. Zoom in to see the difference.}\vspace{-0.5cm}
		\label{fig5}
	\end{figure}

	\subsection{Skip Connection}
	How does skip connections affect the result of SR? What do we lose if we bypass some skip connections? To obtain some insights in this problem, we train PAN with different levels of skip connections on $\times2$ scale (from $128\times128$ to $256\times256$). In Fig.~\ref{fig4}, images on the left are the results using all skip connections. While as we traverse to the right, we see the results layer-by-layer omitting them. We observe that skip connection over coarse spatial resolutions brings high-level semantic consistency such as same pose and similar general hair style while fine-resolution skip connections introduce detail consistency such as expression and wrinkle. Besides, the model with fewer skip connections seems to get harder to converge and tends to produce more artifacts, meaning that these connections indeed make a profound difference to our SR model.

	\subsection{Generalization}
	It is well known that SR of large scale is very challenging endeavor and is prone to produce unrealistic image. But focusing on domain-specific dataset makes it possible for us to hallucinate details appropriately and faithfully,
	thus our method can be easily extended to handle SR problem of different scales. Besides, our method does not need face-specific information such as face landmarks and parsing maps, which makes our method applicable to other object categories. We also try our model ($\times4$) on LSUN cat dataset\cite{LSUN}. 
	
	As illustrated in Fig.~\ref{fig5}, our results largely outperforms other methods perceptually. Though higher variation in dataset results in some artifacts, we notice that these artifacts usually locate in backgrounds, which means that our model apparently know what cat is and how to super-resolve a cat.
	Moreover, thanks to the mems in the training set which contains plenty of texts, some text regions in the background also get good SR results when we super-resolve a cat. So a text-specific SR model may also work using our method.
	
\begin{figure}[t]\vspace{-0.5cm}
		\centering
		\includegraphics[width=\linewidth]{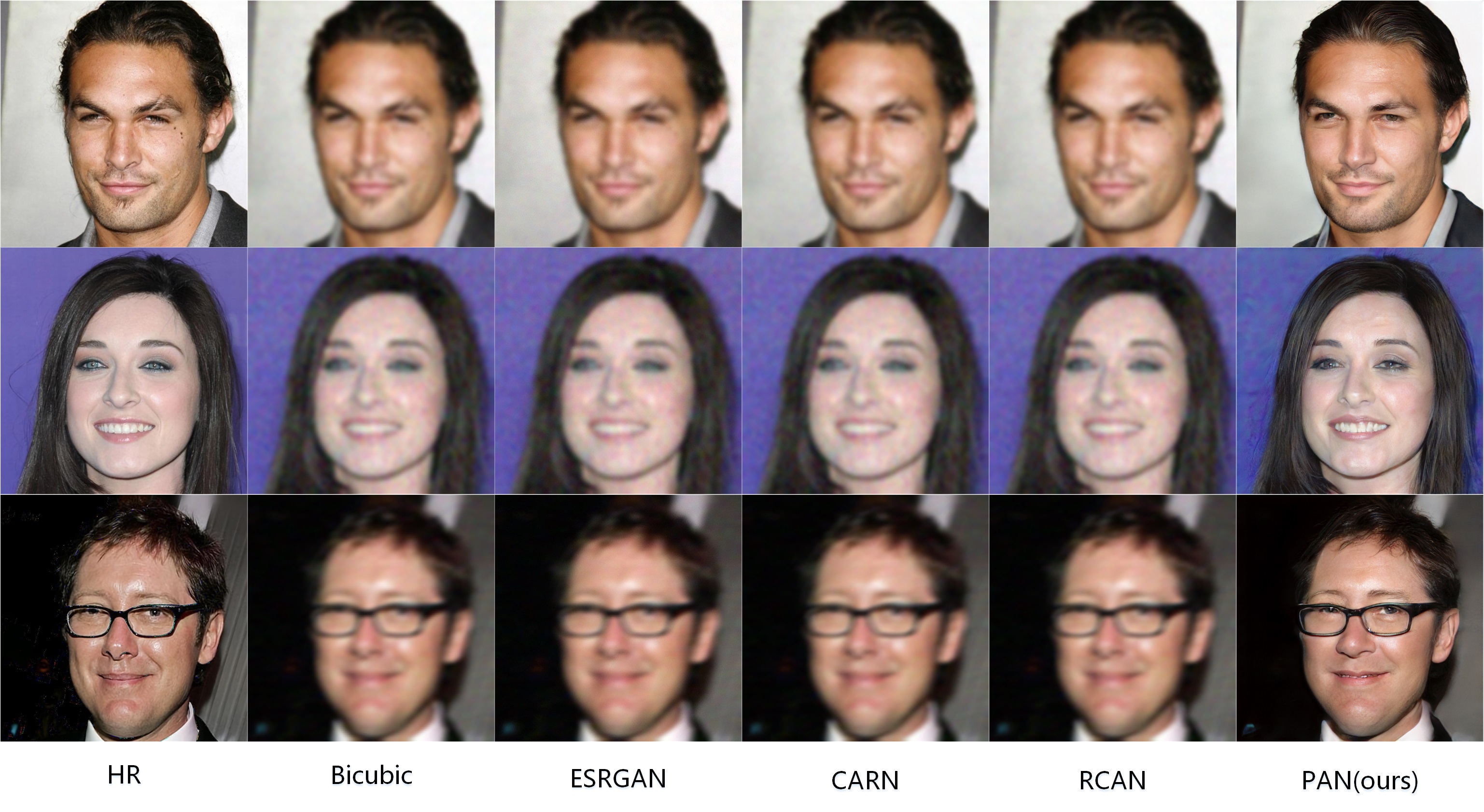}
		\caption{Visual results for different methods with degraded inputs ($\times4$ SR).}\vspace{-0.3cm}
		\label{fig6}
	\end{figure}

	Since real-world low and high-resolution image pairs are not trivially available, in most common strategy for learning super-resolution models, images are first downscaled in order to create corresponding training pairs. As a consequence, however, the resulting low-resolution image is clean and almost noise-free. This often leads to dramatic artifacts when the algorithm is applied to images that come straight from the internet or distinctive cameras. To tackle this, we introduce an online random image degradation to the LR images during training. 
	The image degradation is performed by $\tilde{\vx} = J((\vx \otimes k \otimes b){ \downarrow _s} + \bm n)$, 
	%the following operation: %\vspace{-0.1cm}
	%
	%\begin{equation}
	%\tilde{\vx} = J((\vx \otimes k \otimes b){ \downarrow _s} + \bm n), \vspace{-0.1cm} \label{eq2}
	%\end{equation}
	%
	where $\vx$ is a clean image to be degraded, $k$ and $b$ are the continuous point spread functions caused by camera 
	aperture and handshake, respectively, $\downarrow_s$ is the downsampling, $\bm n$ is noise, $J$ denotes JPEG compression, and $\tilde{\vx}$ is a noisy, blurry, low quality image. By randomly introduce the degradation into input LR images during training, the generalization of our model to low-quality image is significantly improved.
	And notably, compared to the low quality LR image, our SR results have better details and fewer noise, which are traditionally thought to be at odds with each other. The results in Fig.~\ref{fig6} further proves that our model is  robust in practical application.

	\section{Conclusion}
	In this paper, we propose Progressive Adversarial Network (PAN) to produce domain-specific SR image with high-fidelity and high-quality. To circumvent the curse of dimensionality in image super-resolution, we find an alternative to generate high-resolution images instead of using $L_1$/$L_2$ losses. Progressive growing of algorithmic architecture and a partial U-Net are simultaneously applied to achieve reconstruction precision. Adversarial learning with random noise injection in generator is performed to facilitate the high-fidelity and photo-realistic effect of super-resolved images. Extensive experiments are conducted to demonstrate the effectiveness and robustness of PAN. 
	
	The super-resolution images recovered by our algorithm is perceptually accurate and plausible instead of being measured with pixel-wise reconstruction precision, in the sense that the super resolved images are visually perceived identical to ground truth but the tiny details may differ. We call this new fashion of image super-resolution perceptual image super-resolution. 
	
	We also see some promising results for other image-to-image translation problems such as denoising and deblurring. The principle of our algorithm is applicable to these tasks as well. We will study these problems in further work.

	\newpage
	
	% ---- Bibliography ----
	%
	% BibTeX users should specify bibliography style 'splncs04'.
	% References will then be sorted and formatted in the correct style.
	%
	\bibliographystyle{splncs04}
	\bibliography{ref}
\end{document}